# MEASURING URBAN ROAD NETWORK RESILIENCE TO EXTREME EVENTS: AN APPLICATION FOR URBAN FLOODS

**André Borgato Morelli[1]**
**André Luiz Cunha[1]**

**ABSTRACT**
Designing and maintaining resilient transportation systems rely on identifying potential vulnerabilities and inefficiencies before crises occur. However, given the complexity of transportation networks, as well as the diversity of ways in which systems can fail, the problem of assessing the impacts of exceptional phenomena still lacks tools and metrics for analysis. Therefore, this paper aims to present a method for assessing the behavior of road networks during crises in which segments are disabled or unusable for a time. The method proposed was structured around two metrics calculated from the road network and trip distribution: **network continuity** and **efficiency of alternative**. With these metrics, we assess, as an example, the local and global impacts of various flooding scenarios in a case study at São Carlos, a medium-sized Brazilian city in the state of São Paulo. Our findings indicate that pedestrian movements are significantly less impacted than motorized vehicles by floods in the city, which may be due to the car-oriented nature of the avenues closest to rivers, and the shortest average length of pedestrian trips. These findings and metrics set a framework for further research involving other cities and natural phenomena to enrich the understanding of resilience in urban transportation networks.

## 1. INTRODUCTION

Transportation planning has traditionally considered a static road system where traffic does not suffer from the influence of external factors such as environmental catastrophes. This alongside the deterministic conception of nature was associated with the twentieth-century modernist thinking (Ahern 2011). Recently, however, transportation systems have also been considered in terms of adverse situations, to plan for more resilient systems in fault situations (Appert and Chapelon 2007; Chan and Schofer 2016; Martins, Rodrigues da Silva, and Pinto 2019; Westrum 2006).

Resilience is often defined as the capacity of a system to adapt when exposed to adverse situations, avoiding potential losses (Westrum 2006). However, considering the diversity of ways in which a transportation system can fail, with some being more frequent than others depending on the region's demographic and geographical conditions, building a single model to cover all planning for resilient transportation systems becomes a challenge. Thus, different ways to target the problem were devised, some measuring the overall capacity of systems to absorb impacts (Ip and Wang 2011; Leu, Abbass, and Curtis 2010; Morelli and Cunha 2019b) and others focusing on specific events such as an oil supply crisis (Martins et al. 2019; Newman, Beatley, and Boyer 2009), hurricanes (Beheshtian et al. 2018; Chan and Schofer 2016; Litman 2005) or floods (Gil and Steinbach 2008; Morelli and Cunha 2019a). However, there is little consensus between the various types of analysis, making the integration of models a highly complex process. This reality highlights the need for more comprehensive methods that can be easily integrated with other established urban planning methods.

---

[1] Department of Transportation Engineering, São Carlos School of Engineering, University of São Paulo. Av Trabalhador Sãocarlense, 400, São Carlos, SP, Brazil. Corresponding author: andre.morelli@usp.br

## 2. LITERATURE REVIEW

Studies in resilience of urban transportation generally assess the ability of the network to absorb atypical events. More generalist models aim to measure network redundancy, robustness, or vulnerability, without previously defining the problem to be addressed by the city in a crisis (Berche et al. 2009; Ip and Wang 2011; Leu et al. 2010; Morelli and Cunha 2019b). Several works with this general nature use graph theory to understand system fragility, like Appert and Chapelon (2007) who studied the impact of removing specific edges on Montpellier's main road network in France; a similar analysis, conducted by Rodríguez-Núñez and García-Palomares (2014), focused on public transportation infrastructure, removing connections from the Madrid subway system, keeping track of the routes affected by the action in an attempt to define the importance order of the systems' links. Berche *et al.* (2009) analyzed the effect of systematic attacks with graph theory parameters to transit networks in 14 cities around the world, in which the author defines as "attack" the removal of connections in the network. Ganin *et al.* (2017) focused on random network degradation, this time measuring the changes in the total delay in 40 cities in the United States. In Brazil, Morelli and Cunha (2019b) applied systematic attacks on the networks of the 306 largest Brazilian cities, measuring system continuity based on the number of routes that remain viable after an attack. Doing this, they determined the strategy that imposes faster degradation and found that edges with high betweenness centrality tend to be more vulnerable.

On the other hand, some studies focus on more specific problems, such as oil supply crises (Leung, Burke, and Cui 2018; Martins et al. 2019; Newman et al. 2009) and natural disasters (Gil and Steinbach 2008; Litman 2005; Lu, Peng, and Zhang 2014; Morelli and Cunha 2019a). Martins, Rodrigues da Silva e Pinto (2019) look at changes in travel patterns by analyzing the ability of users to change from fuel-dependent modes to active ones looking at traveled distances in the network. Chan and Schofer (2016) analyzed the decay of passenger numbers in two large urban rail systems during hurricanes. Some studies assess the impact of flooding on networks (Gil and Steinbach 2008; Morelli and Cunha 2019a) through a geometric standpoint, not taking into account the travel distribution on the studied regions.

In the cases above, works focusing on travel distribution tend to have little focus on the structure of the transportation network, while works focusing on connectivity and structure, tend to relegate travel behavior to the background, making both methods difficult to integrate with an analysis that can take into account the accumulated effects of both phenomena.

## 3. PROPOSED METHOD

The strategy proposed to assess the impacts of link deactivation consists of calculating two different indices: one to measure continuity – a measure of redundancy – and another to measure the efficiency of alternative paths. The method is summarized in the flowchart in Figure 1.

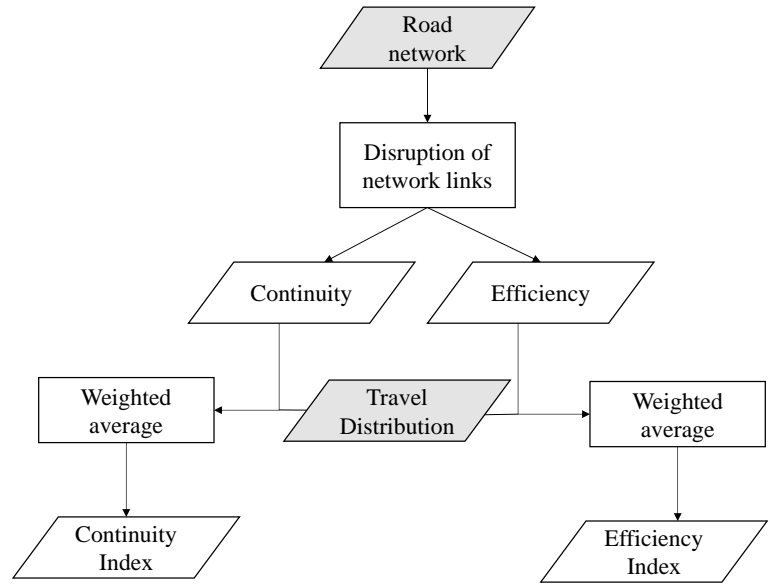

**Figure 1: Flowchart of the proposed method.**

Only two datasets are needed as input for these steps: a city's road network and travel distribution between traffic zones (Origin/Destination or OD matrix), both marked in light grey in Figure 1. These types of datasets are common in transportation planning practice, facilitating the incorporation of the method by stakeholders in decision making. The analysis was conducted through libraries of graph processing and geoanalysis from Python programing language.

### 3.1. Network disruption

The disruption of a network occurs for different reasons depending on the impact studied. However, the effects on network structures can be reduced to a simple phenomenon: an event acts to limit or prevent movement on a road or a series of them. In this paper we targeted events that prevent movement along links. Figure 2 shows possible impacts on a route when the event may:

- Not affect the route (Figure 2(a));
- Divert a route (Figure 2(b)), increasing its length;
- Affect the route and all its alternatives (Figure 2(c)) rendering the route impossible;

In these cases, we assume the user will always follow the shortest path connecting two points. Therefore, a damaging event on a road network can never decrease route lengths.

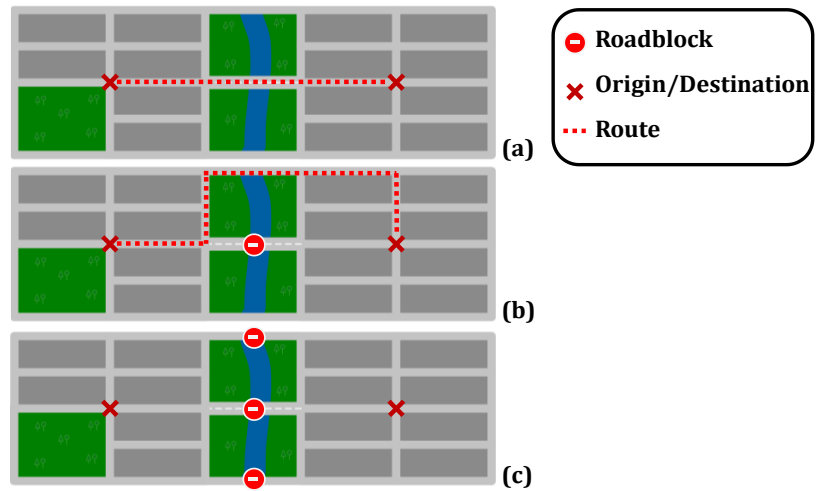

Figure 2: Types of impact for a single route.

An urban road system will be denoted as a graph to simulate events where network disruption occurs. In this mathematical structure, road segments are represented by edges and intersections are represented by nodes. This representation is commonly referred to as the primal representation of the road network (Porta, Crucitti, and Latora 2006). In this configuration, a trip can be generated from any node in the system to any other, traversing the edges of the graph. Figure 3 shows a fictional urban region represented by a graph and a route within this network (in red).

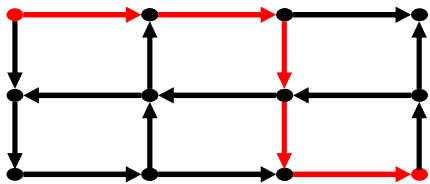

Figure 3: Graph of a fictional urban network with a route (red).

The blockage of movement along a road segment can be achieved by removal of the referred edge from the graph, so the minimum paths generated in the disrupted network ignore the blocked road. However, it is necessary to differentiate between the impacts for modes that follow the direction of the road (bicycles, cars, motorcycles, and motor vehicles in general) and pedestrian movements, which can traverse the network in either direction of the links. Apart from this, due to municipal restrictions, some roads may also limit or not allow traffic to some modes of transportation. In this case, the networks defined for the various types of movement are distinct, requiring a split-mode impact assessment. In this paper we evaluate the impacts on pedestrians (with an undirected graph) and motorized vehicles (with a directed graph).

The OSMnx library (Boeing, 2017) in Python language was used for extracting graph networks. This library extract databases directly from the OpenStreetMap platform, which differentiates walking and driving networks. This distinction is important because both distances traveled, and accessibility depends on how agents interact with the network. For example, a pedestrian's path may be shorter than that of a motorized vehicle, as shown in Figure 4(a), or a path that is possible for a pedestrian may not be available for a motorized vehicle after the deactivation of a link, as can be seen in Figure 4(b).

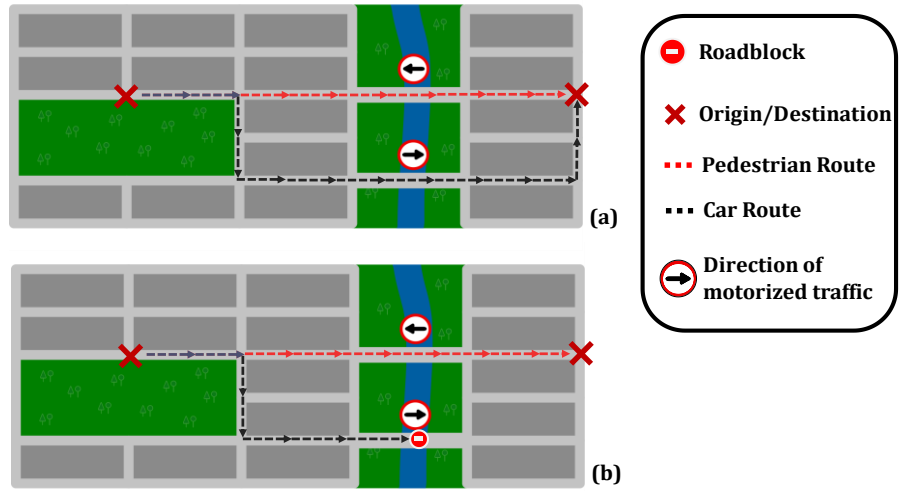

Figure 4: Pedestrian and motorized vehicle travel examples - (a) A pedestrian path could be shorter than a car route; (b) some events may block movements of one mode of transportation but not all.

### 3.2. Impact between traffic zones

The impact of network degradation depends mainly on how travel routes from a OD matrix are allocated. The average length of trips can sometimes be estimated by the route between centroids of traffic zones, as in Martins *et al.* (2019) and represented in Figure 5 (a), but this approach would lead to poor results when evaluating changes happening in a network structure, as the metrics become solely dependent on links used for the path between centroids. In this case, to measure the impact of network degradation it is necessary to consider a more diverse set of routes (Figure 5 (b)). Thus, to tackle this problem we propose the use of a random sample of routes that start in zone A and end in zone B to represent the impact on the OD pair A/B. This sample has to be large enough to capture the properties of the pair, but not too large to limit the computational load. For the case study in this paper, as we evaluate a mid-sized city, we decided to consider all possible pairs between zones, but larger urban areas may require down sampling. The paths were calculated through Dijkstra's algorithm (Dijkstra, 1959) using the iGraph library (Csardi and Nepusz, 2006).

Thereby, we can obtain the interconnectedness and the average route length between each OD pair, so we can define two metrics: **network continuity** and **efficiency of alternative**, which are explained in the sections below.

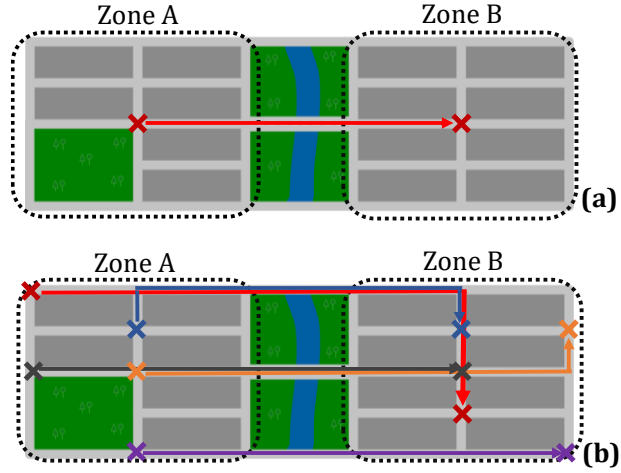

**Figure 5:** Travel analyzes – (a) centroid approach; (b) random sampling approach.

*3.4.1. Network continuity*

In short, a network has perfect continuity if all nodes can be reached in the graph from any other node. On the other hand, if an exceptional event, such as a flood, isolates two or more blocks in a city, preventing traffic between them, the city loses continuity. Hence, continuity between two traffic zones is proposed to be the ratio of node-to-node paths of the network graph which remain valid between zones after an impact, so that:

$$\mathbf{C(A,B)} = \sum_{s \in A} \sum_{t \in B} \frac{V(s,t)}{N_A \cdot N_B} \qquad (1)$$

Where:   $A$ e $B$: Traffic zones in the network;
$s$: Node belonging to zone A;
$t$: Node belonging to zone B;
$N_K$: Number of nodes in zone K;
$V(s,t)$: Validity function. Assumes 1 if the route from $s$ to $t$ exists and 0 if not.

Equation 3 considers only one type of network, represented by a single graph. However, as discussed earlier, different modes of transportation have movements of different nature, requiring different types of graph structures. In this case, continuity can be measured for each mode of transportation separately.

*3.4.2. Efficiency of alternative*

Route efficiency can be inferred from the average minimum distances between traffic zones. When a network element is removed, routes that previously depended on that element must be diverted, generally increasing the total path length. A network is efficient at absorbing an impact when it offers similar routes connecting nodes, requiring no large detours and leading the average increase in distances to be minimal or zero.

The distance between two points can only be measured for valid routes in the system. In this case, it is usual to consider the proximity between points as a metric, instead of the distance. The proximity of two points is the inverse of the distance between then, as in:

$$P(s,t) = \frac{1}{d(s,t)} \quad (2)$$

Where:   $P(s,t)$: Proximity between **s** e **t**;
$d(s,t)$: Distance between **s** and **t**.

Likewise, two points that are not reachable from each other can be defined as having zero proximity, making it unnecessary to disregard any pair of points in the calculation. Therefore, route efficiency between two traffic zones A and B can be defined as the average route proximity like:

$$\eta(A,B) = \sum_{s \in A} \sum_{t \in B} \frac{P(s,t)}{N_A \cdot N_B} \quad (3)$$

It should be noted that, particularly in motorized vehicle routes, the efficiency from zone *A* to zone *B* is not necessarily equal to the efficiency from *B* to *A* and the existence of a route from nodes *s* to *t* does not imply the existence of a route from *t* to *s* given the directed nature of the graph. Thus, both the efficiency and continuity functions are generally non-symmetrical.

*3.4.3. Relative impact in traffic zones*
To measure relative impact, we weighed the total number of trips taking place between traffic zones on the continuity and efficiency metrics. This assumption captures the fact that impacts on movements between low demand OD pairs are less important than impacts on movements of high demand OD pairs. Besides, we consider the effects of impacts on different modes of transportation as a proportion of trips dependent on each mode. Therefore, the impact of a phenomenon in a traffic zone is given by the weighted average of the impacts generated in trips with origin or destination in the traffic zone, thus:

$$C(A) = \sum_{K \in G} \frac{[C(A,K) \cdot T(A,K) + C(K,A) \cdot T(K,A)]}{T(A,K) + T(K,A)} \quad (4)$$

$$\eta(A) = \sum_{K \in G} \frac{[\eta(A,K) \cdot T(A,K) + \eta(K,A) \cdot T(K,A)]}{T(A,K) + T(K,A)} \quad (5)$$

Where:   *G*: Network graph;
*K*: Traffic zone from graph *G*;
*A*: Reference traffic zone;
$T(Z1, Z2)$: Total trips between zones *Z1* e *Z2* using the mode *m*;

$C_m(Z1, Z2)$: Continuity between zones $Z1$ and $Z2$ for mode $m$;
$\eta_m(Z1, Z2)$: Efficiency between zones $Z1$ and $Z2$ for mode $m$.

And finally, when assessing resilience generally there is no intention to evaluate the continuity or efficiency of a network in absolute terms during an exceptional event, but rather the relative reduction in these parameters during the event. To measure this, efficiency and continuity are defined for an unchanged state ($\eta_0$ and $C_0$) and for the post-event scenario in a disturbed state ($\eta_f$ and $C_f$), with the relative index for an event being calculated as a proportion. Thus, the continuity and efficiency indices may be defined, respectively, as:

$$I_C(A) = \frac{C_f(A)}{C_0(A)} \qquad (6)$$

$$I_E(A) = \frac{\eta_f(A)}{\eta_0(A)} \qquad (7)$$

Where:  $C_0$ ; $\eta_0$: Conditions before impact;
$C_f$ ; $\eta_f$: Conditions after impact.

## 4. APPLICATION AND DISCUSSION

To test the method effectiveness, the impacts of flooding scenarios in São Carlos, São Paulo, Brazil, as defined by Morelli and Cunha (2019a), are evaluated in this session. The distribution of trips in the city is given by the OD survey conducted in 2007/2008 for the municipality, with the division of the territory into 41 traffic zones. In the universe of interviewees, an average of just over 6,000 trips per day were recorded, distributed among walking, cycling, individual motorized, bus, and other modes. For this paper, only walking, cycling, and individual motorized trips are considered. Also, we consider only the most important trips, disregarding those with leisure motives. As for the road network, the topological data was obtained from the open-access collaborative mapping platform OpenStreetMap while the topographical data of the nodes – for the flooding scenarios – was obtained from the Google Maps API (Google Maps API, 2019).

### 4.1. Flood scenarios

The Master Plan of the city, created in 2005 and still in effect, delineates the areas at greater risk of flooding at the time. Since then and going forward, the effects of climate change, urbanization, and poor drainage may cause the expansion of flood risk areas. In this case study, we aim to simulate an increase in flooding intensity by gradually expanding the risk zones through fixed rises in water level. We define the baseline scenario as a flood affecting the risk areas defined in the Master Plan. As the intensity increases and the water level ($\Delta y$) go up, areas at higher elevation begin to flood as well, expanding blockages created by floods. The blockage of traffic by flooding was simulated with the removal of all edges in the graph that was connected to at least one flooded node. We raised the water level in steps of 10 cm from the baseline up until 1.5 m. Some of these flooding scenarios can be seen in Figure 6.

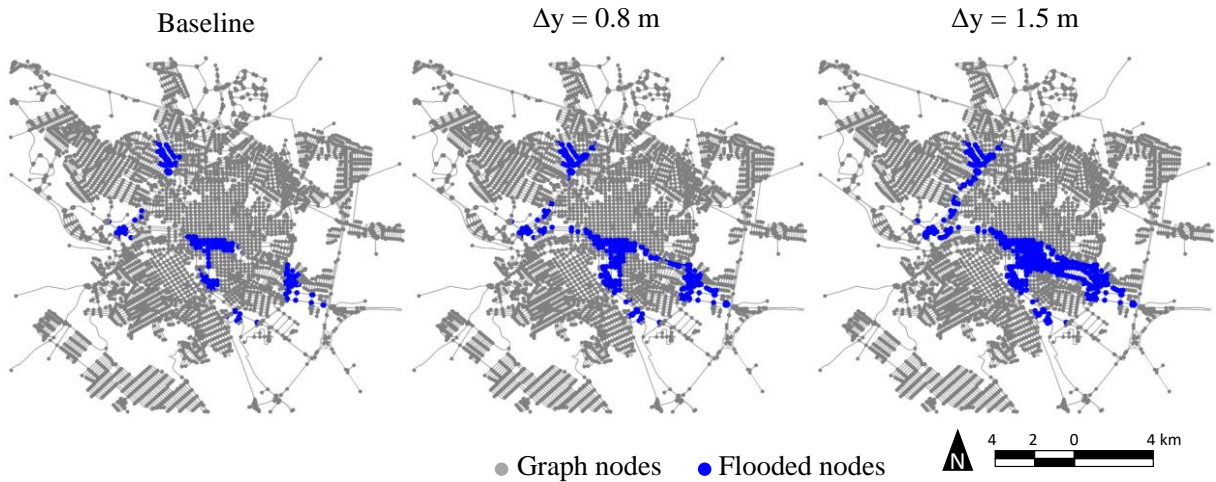

● Graph nodes ● Flooded nodes

**Figure 6: Flooding scenarios in the city of São Carlos, São Paulo State, Brazil.**

In each scenario, the flooded edges are removed from the graph, and the impact of the flooding events in continuity and efficiency is measured using Equations 6 and 7, considering the unaltered graph as the initial state and the flooded graph as the final. These metrics were applied to two types of graph. The first – directed – is relative to motorized vehicles and bikes; and the second – undirected – is relative to pedestrians. The relative results for bicycles and motorized vehicle flows are represented in Figure 7, while Figure 8 shows the results for pedestrians. We observed that, in general, the southern regions of the city are more impacted than the northern regions, rendering the population in these areas more vulnerable to extreme flood events. This occurs in both motorized and pedestrian movements similarly, as there are no expressive differences between the distribution on pedestrian and motorized impacts. On the other hand, the overall impact on the system tends to show a large difference between modes, as depicted in Figure 9, which contains the curve for motorized vehicles and bicycles (Figure 9 (a)), pedestrians (Figure 9 (b)) and the whole system (Figure 9 (c)). Pedestrian movements suffer lower impacts on average if compared to motorized vehicles and bicycles. We hypothesize that this is due to the nature of the movements in question, as the marginal avenues of the city – which are more prone to floods – are largely car-oriented and very important to the city road network for relatively long-distance trips, as it connects the residential suburbs to the commercial center. On the contrary, pedestrians tend to focus their movements on the more densely populated core of the city and to make relatively short trips. This way, although the network is impacted almost the same way for pedestrians and motorized vehicles, the nature of the trips is responsible for the largest difference in the resilience of the trip.

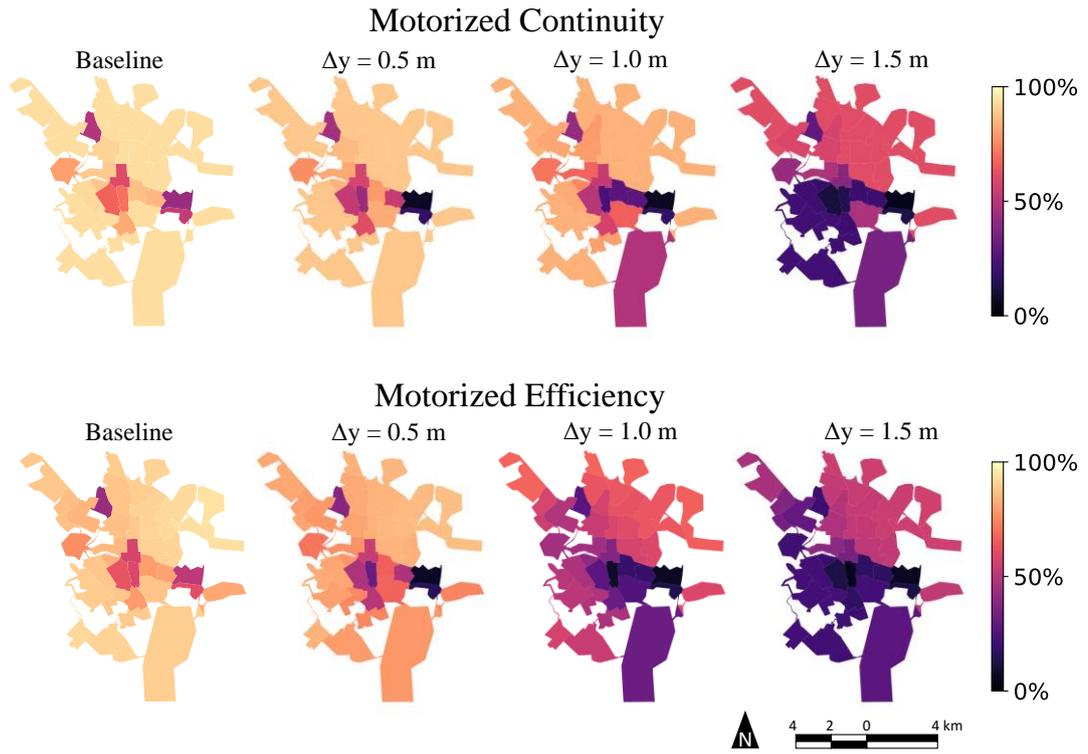

**Figure 7:** Motorized vehicle metrics. Darker shades indicate that a larger proportion of paths from/to the traffic zone are no longer possible after the flood.

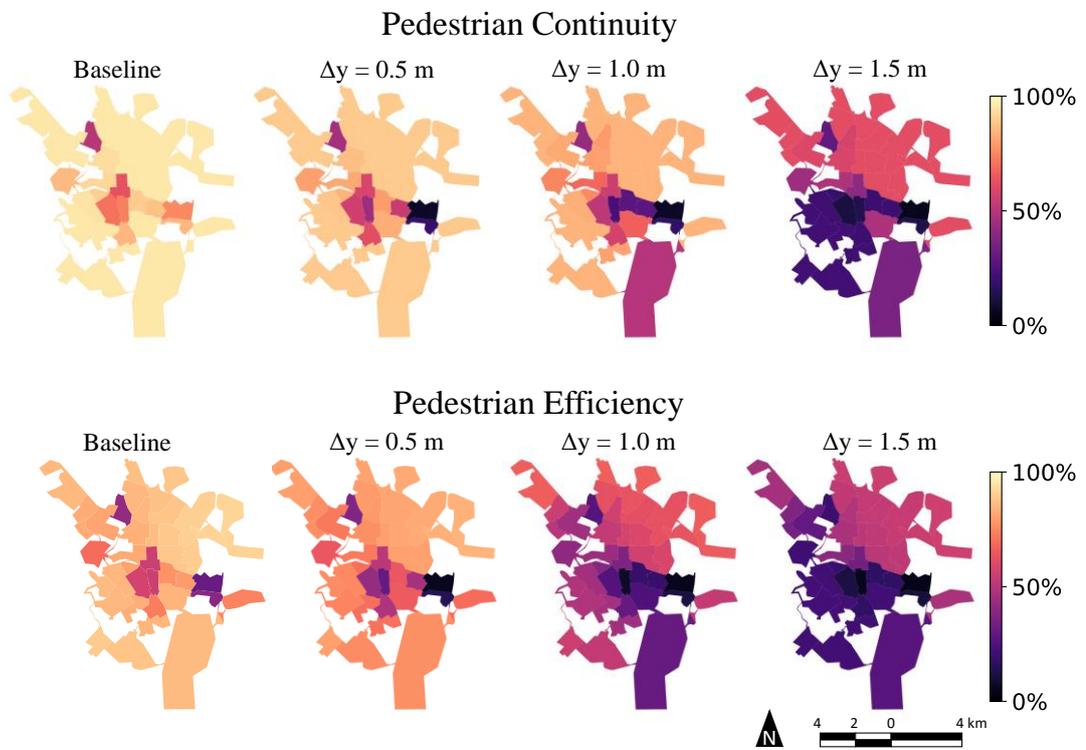

**Figure 8:** Pedestrian metrics. Darker shades indicate a larger increase in average shortest route lengths from/to the traffic zone.

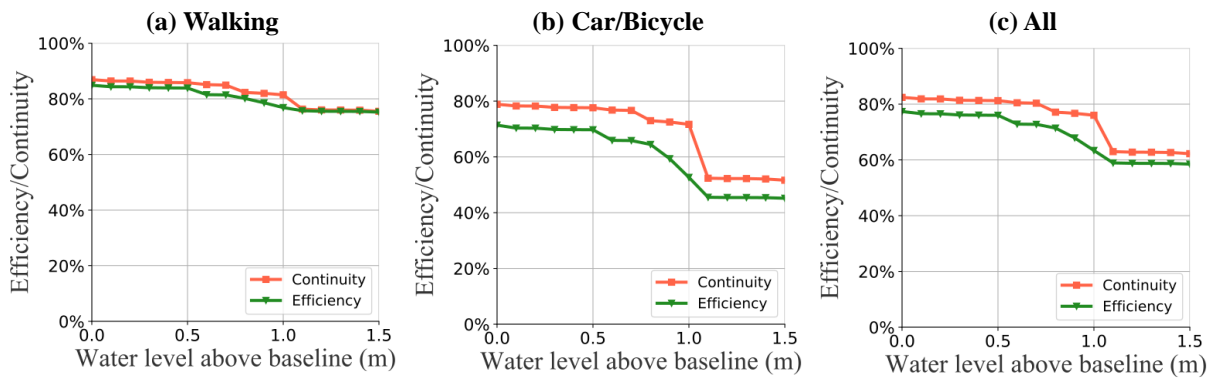
**Figure 9: Results for (a) Pedestrians, (b) bicycles and motorized vehicles, and (c) overall**

Moreover, floods that exceed 1.0 m from the baseline impose major decreases in the metrics for the network, with 55% of car routes being blocked and the efficiency of paths falling to 50%, which means the routes during the events are twice as long on average. For pedestrians, on the other hand, there are no large dips in the curves. This is another evidence that motorized vehicles are more vulnerable to blockages, particularly in arterial roads, which tend to create bottlenecks in the system.

## 5. CONCLUSION

We proposed a novel method for measuring the resilience of transportation to extreme events in urban road networks based on travel distribution. The databases required for the analysis are the road network and the OD matrix, both already common in the practice of urban planning, which facilitates the integration of this method in decision-making processes and favors the creation of a consensus on the techniques used for planning resilience in transportation networks.

Two metrics are proposed: **network continuity** and **efficiency of alternative**. Continuity measures the proportion of routes blocked by a network disruption event. This metric can indicate how parts of a city may become isolated during an exceptional event, making these regions inaccessible to users, residents, and emergency vehicles. The efficiency of alternative is indicative of the average increase in route lengths, which impact on fuel consumption and time loss on motorized vehicles' trips. However, they impact more directly active modes of transportation, like bicycles and pedestrians, which are forced to traverse larger distances to get to their destinations and possibly render the trip impossible if the maximum walking or cycling distance for the user is exceeded.

As a case study, these metrics were applied to measure the impact of flooding in the city of São Carlos, a mid-sized city in the state of São Paulo in Brazil. It was observed that extreme flooding events add significant stress to the network, with the southern regions of the city accounting for most of the impact. The overall impact for pedestrians is significantly smaller than that of motorized vehicles, probably due to the average length of the trips being larger for vehicles, and to the car-oriented nature of the avenues closest to rivers in this city, as happens in many other cities in the world. This is an indicator that more densely packed cities, with shortest trips overall, tend to be more resilient to these extreme events, since longer paths have a higher probability of depending on a street segment affected by the flood, which is to be tested in further research.

The case study is an example of the applicability of these metrics in other research to evaluate in a relatively simple way the impacts of extreme events of various natures in different cities, setting a framework for further research on transport network's resilience. Also, as the data needed for this analysis are relatively common in mid- to large-sized cities, these metrics may prove to be good guides to decision-makers and city planners to create policy on transportation systems.

**Funding**
This study was financed in part by the Coordenação de Aperfeiçoamento de Pessoal de Nível Superior - Brasil (CAPES) - Finance Code 001.

André Borgato Morelli (andrebmorelli@gmail.com)
Prof. Dr. André Luiz Cunha (alcunha@usp.br)
Department of Transportation Engineering, São Carlos School of Engineering, University of São Paulo.
Av Trabalhador Sãocarlense, 400, São Carlos, SP, Brazil.